# Electromagnetic Tunneling of Obliquely-Incident Waves through a Single-Negative Slab Paired with a Double-Positive Uniaxial Slab

**Giuseppe Castaldi,[1] Vincenzo Galdi,[1,*] Andrea Alù,[2] and Nader Engheta[3]**

[1]*CNR-SPIN and Waves Group, Department of Engineering, University of Sannio, I-82100, Benevento, Italy*

[2]*Department of Electrical and Computer Engineering, The University of Texas at Austin, Austin, TX 78712, USA*

[3]*Department of Electrical and Systems Engineering, University of Pennsylvania, Philadelphia, PA 19104, USA*

[*]*Corresponding author: vgaldi@unisannio.it*

We show that, under appropriate oblique-incidence and polarization conditions, the inherent opaqueness of a homogeneous, isotropic *single-negative* slab may be perfectly compensated (in the ideal lossless case) by a homogenous, anisotropic (uniaxial) *double-positive* slab, so that *complete tunneling* (with total transmission and zero phase delay) occurs. We present an analytical and numerical study aimed at deriving the basic design rules, elucidating the underlying physical mechanisms, and exploring the role of the various involved parameters.

# 1. Introduction and Background

*Single-negative* (SNG) materials, characterized by only one negative (real part) constitutive parameter, are essentially *opaque* to the electromagnetic (EM) radiation, in view of the dominant *imaginary* character of the propagation constant, even in the ideal lossless case. However, they can give rise to very intriguing, and somehow counterintuitive, field effects when inserted in heterostructures under proper matching conditions. For instance, Fredkin and Ron [1] observed that a layered material composed of alternating *epsilon-negative* (ENG) and *mu-negative* (MNG) layers, in spite of the inherent opaqueness of its constituents, was capable of supporting *propagating* modes effectively exhibiting either a *negative*- (see also [2]) or *positive*-refractive-index character. Alù and Engheta [3] focused instead on homogenous, isotropic ENG-MNG bi-layers, and showed that resonant tunneling phenomena (with total transmission and zero phase-delay) might occur under suitable matching conditions. Some of their results may also be interpreted in the more general framework of *complementary media* introduced by Pendry and Ramakrishna [4], which also allows straightforward extension to anisotropic, inhomogeneous configurations.

Building on the above results, further extensions and generalizations have been proposed in [5-21], including one-dimensional photonic crystals [5,6,9,10,12], non-contiguous layers [15], "non-conjugated" pairs [17], transformation-optics-inspired configurations [19], as well as general heterostructures containing impedance-mismatched metamaterial layers [13] and, more specifically, SNG layers paired with *double-negative* (i.e., negative permittivity and permeability – DNG) [11,14] or *double-positive* (i.e., positive permittivity and permeability – DPS) [7,8,16,18,20,21] layers.

Especially relevant for the present investigation are the SNG-DPS configurations in [7,8,16,18,20,21]. While it can readily be proved analytically [7] that, for *normal incidence*, a *single* (homogeneous, isotropic) DPS slab *cannot* perfectly compensate the opaqueness of a SNG slab, it was shown in [7,8,18,21] that *complete tunneling* may actually occur through an ENG layer *symmetrically sandwiched* between high-permittivity DPS layers (see also the related theoretical and numerical study in [22]). Such results were also demonstrated experimentally at microwave frequencies by synthesizing the required metamaterials via resonant (H-shaped, mesh, split-ring) metallic inclusions [7,8,18]. In [20], we extended the above results to *asymmetrical* tri-layers composed of an ENG slab paired (at one side only) with a bi-layer of homogeneous DPS materials. For an assigned frequency and normally-incident illumination, we derived analytically the design rules for such DPS bi-layer to compensate the opaqueness of a given ENG slab. Moreover, we showed that, under the above conditions, the DPS bi-layer would *effectively mimic* (in terms of wave impedance and reflection properties) an equivalent "matched" (according to [3]) MNG slab, over a moderately wide (~10%) bandwidth. Considering the relative challenge in realizing magnetic metamaterials with negative permeability, particularly at higher frequencies, it is remarkable that a proper combination of DPS slabs may effectively act as an MNG layer.

Interestingly, it was shown in [16] that *zero reflection* may actually occur for an ENG-DPS bi-layer, under *obliquely-incident* transversely-magnetic (TM) polarized illumination. However, the study in [16] was limited to the bare mathematical derivation of the reflectionless condition. In this paper, we study in detail the hitherto unexplored physical mechanisms underlying this very intriguing phenomenon and their possible implications. In this framework, we consider a more general configuration featuring a homogeneous, isotropic SNG slab paired with a *uniaxially anisotropic* DPS slab, which provides an additional tuning parameter. Unlike the Fabry-Perot-type resonant phenomena observed in [7,8,18,20,21] (characterized by standing waves in the DPS layers, and nonzero phase-delay), the resonant phenomena in the proposed configuration are mediated by the excitation of *localized surface modes* at the SNG-DPS interface, are characterized by zero phase-delay, and depend on the slab thickness *ratio* (rather than *sum*), in a much closer analogy with what observed in the matched ENG-MNG bi-layers [3]. This allows to establish a more direct and physically-incisive analogy between the uniaxial DPS slab and a matched (homogeneous, isotropic) SNG slab, which may turn out useful to simplify the realization of some of equivalent magnetic effects at high frequencies.

Accordingly, the rest of the paper is laid out as follows. In Sec. 2, we outline the problem geometry and formulation. In Sec. 3, we derive the main analytical results, starting with the resonant tunneling conditions, and proceeding with the analogy between the uniaxial DPS slab and a matched SNG slab. In Sec. 4, we present and discuss some representative numerical examples, and investigate the sensitivity of the tunneling phenomena to frequency, polarization and incidence direction of the illumination, as well as the effects of the unavoidable material dispersion and losses. Some brief concluding remarks follow in Sec. 5.

## 2. Problem Geometry and Statement

Without loss of generality, the two-dimensional configuration under study, illustrated in the Cartesian $(x, y, z)$ reference coordinate system of Fig. 1, comprises a homogeneous, isotropic ENG slab of thickness $d_1$ and relative permittivity $\varepsilon_1$ (with $\mathrm{Re}(\varepsilon_1) < 0$), paired with a homogeneous, *generally anisotropic* (uniaxial) DPS slab of thickness $d_2$ and relative permittivity tensor

$$\underline{\underline{\varepsilon_2}} = \begin{bmatrix} \varepsilon_{2\perp} & 0 & 0 \\ 0 & \varepsilon_{2\|} & 0 \\ 0 & 0 & \varepsilon_{2\|} \end{bmatrix}, \quad \mathrm{Re}\left(\varepsilon_{2\perp}\right) > 0, \mathrm{Re}\left(\varepsilon_{2\|}\right) > 0. \tag{1}$$

Both slabs are assumed as *nonmagnetic* (i.e., $\mu_1 = \mu_2 = 1$), infinitely long in the *y*- and *z*-direction, and immersed in vacuum. We assume time-harmonic $\left(\exp\left(-i\omega t\right)\right)$, unit-amplitude, obliquely- incident (with angle $\theta_i$, cf. Fig. 1) TM-polarized plane-wave illumination, with *z*-directed magnetic field

$$H_z^i\left(x, y\right) = \exp\left[ik\left(x\cos\theta_i + y\sin\theta_i\right)\right], \tag{2}$$

where $k = \omega/c = 2\pi/\lambda$ denotes the vacuum wavenumber, and $c$ and $\lambda$ the corresponding speed of light and wavelength, respectively. In what follows, we outline the general analytical solution of the problem, and derive the conditions for *total transmission*.

## 3. Analytical Derivations

### *A. Generalities*

In the ideal *lossless* case, the magnetic field expression can be written as

$$H_z(x,y)=\begin{cases} H_z^i + B_0 \exp\left[ik\left(-x\cos\theta_i + y\sin\theta_i\right)\right], & x<-d_1, \\ \exp\left(iky\sin\theta_i\right)\left[A_1\cosh\left(\alpha_1 x\right)+B_1\sinh\left(\alpha_1 x\right)\right], & -d_1<x<0, \\ \exp\left(iky\sin\theta_i\right)\left[A_2\cosh\left(\alpha_2 x\right)+B_2\sinh\left(\alpha_2 x\right)\right], & 0<x<d_2, \\ A_3\exp\left[ik\left(x\cos\theta_i + y\sin\theta_i\right)\right], & x>d_2, \end{cases} \tag{3}$$

where the phase-matching conditions (conservation of the transverse wavenumber) at the interfaces $x=-d_1,0,d_2$ and the radiation condition are already enforced, and

$$\alpha_1 = k\sqrt{\left|\varepsilon_1\right| + \sin^2\theta_i}, \tag{4}$$

$$\alpha_2 = k\sqrt{\varepsilon_{2\parallel}\left(\frac{\sin^2\theta_i}{\varepsilon_{2\perp}}-1\right)}, \quad \sin^2\theta_i > \varepsilon_{2\perp}, \tag{5}$$

denote the (real, positive) attenuation constants in the ENG and uniaxial DPS slabs, respectively. The inequality in (5) ensures that the uniaxial DPS slab operates *below cutoff*, which is instrumental in the following developments. The six unknown expansion coefficients $B_0, A_1, B_1, A_2, B_2, A_3$ in (3) may be computed by enforcing the tangential-field continuity at the interfaces $x=-d_1,0,d_2$, with the electric field following from (3) and the relevant Maxwell's curl equation.

## B. Conditions for Total Transmission

The analytical expressions of the expansion coefficients in (3) are not reported here for brevity. Instead, we focus on the coefficient $B_0$, which plays the role of the reflection coefficient. Accordingly, the total-transmission resonant condition is derived by zeroing its numerator (provided the denominator is nonzero), viz.,

$$
\begin{aligned}
&\varepsilon_{2||}\alpha_2\left(\alpha_1^2+\varepsilon_1^2 k^2\cos^2\theta_i\right)\tanh\left(\alpha_1 d_1\right)\\
&+\varepsilon_1\alpha_1\left(\alpha_2^2+\varepsilon_{2||}^2 k^2\cos^2\theta_i\right)\tanh\left(\alpha_2 d_2\right)\\
&+ik\cos\theta_i\left(\varepsilon_1^2\alpha_2^2-\varepsilon_{2||}^2\alpha_1^2\right)\tanh\left(\alpha_1 d_1\right)\tanh\left(\alpha_2 d_2\right)=0.
\end{aligned}
\tag{6}
$$

Zeroing the imaginary part of (6), and recalling (4) and (5), we obtain

$$
\varepsilon_{2\perp}=\frac{\varepsilon_1^2\sin^2\theta_i}{\varepsilon_{2||}\left(\left|\varepsilon_1\right|+\sin^2\theta_i\right)+\varepsilon_1^2}, \tag{7}
$$

which automatically satisfies the inequality (cutoff condition) in (5). Substituting (7) in (6), and zeroing the remaining real part yields the second condition:

$$
\frac{\varepsilon_{2||}^2\alpha_1\left(\alpha_1^2+\varepsilon_1^2 k^2\cos^2\theta_i\right)}{\left|\varepsilon_1\right|\cosh\left(\dfrac{\alpha_1 d_1}{\varepsilon_1}\right)\cosh\left(\dfrac{\varepsilon_{2||}\alpha_1 d_1 d_2}{\varepsilon_1}\right)}\sinh\left[\alpha_1\left(d_1-d_2\frac{\varepsilon_{2||}}{\left|\varepsilon_1\right|}\right)\right]=0, \tag{8}
$$

whose general solution is:

$$
\left|\varepsilon_1\right| d_1=\varepsilon_{2||} d_2. \tag{9}
$$

Thus, for a given ENG slab with parameters $\varepsilon_1, d_1$, and for a given incidence angle $\theta_i$, the conditions in (7) and (9) yield *an infinity* of possible solutions [23] for total transmission, in terms of the three remaining parameters $\varepsilon_{2\perp}, \varepsilon_{2||}, d_1/d_2$. For the case of *isotropic* DPS slab $\left(\varepsilon_{2\perp}=\varepsilon_{2||}\right)$, (7) and (9) reduce to the results in [16].

Note that the seemingly possible (trivial) solutions of (8) featuring $\varepsilon_{2||}=0$ or $\alpha_1=0$ are inconsistent with the previous assumptions used to derive (8). Moreover, from (7), it is readily understood that, approaching normal incidence (i.e., $\theta_i \to 0$), *extreme* parameter values (i.e., $\varepsilon_{2\perp} \to 0$) are required in order to achieve complete tunneling.

A few other general considerations are in order. First, the total-transmission conditions in (7) and (9) do not depend *explicitly* on the frequency, although an implicit frequency dependence is unavoidable in view of the inherent material dispersion in ENG materials. Also, they do not depend on the bi-layer *total* thickness, but rather on the *ratio* $d_1/d_2$, thereby implying that the layers may be made, in principle, *arbitrarily thin*. The additional permittivity parameter available in our anisotropic configuration allows more flexibility in the choice of $d_1/d_2$, which is instead bounded for the isotropic case [16]. This may be particularly useful for strongly opaque ENG layers (i.e., $|\varepsilon_1| \gg 1$) and nearly-normal incidence. Next, it can be shown that, under total-transmission conditions, the expansion coefficient $A_3$ in (3) reduces to

$$A_3 = \exp\left[-ik\cos\theta_i\left(d_1 + d_2\right)\right], \tag{10}$$

so that the total phase-delay accumulated through the bi-layer is zero. Moreover, looking at the expression of the field intensity in the bi-layer under total-transmission conditions,

$$\left|H_z\left(x\right)\right|^2 = \begin{cases} \left\{\cosh^2\left[\alpha_1\left(x+d_1\right)\right] + \dfrac{\varepsilon_1^2 k^2\cos^2\theta_i}{\alpha_1^2}\sinh^2\left[\alpha_1\left(x+d_1\right)\right]\right\}, & -d_1 < x < 0, \\ \left\{\cosh^2\left[\dfrac{\alpha_1\varepsilon_{2\|}\left(x-d_2\right)}{\varepsilon_1}\right] + \dfrac{\varepsilon_1^2 k^2\cos^2\theta_i}{\alpha_1^2}\sinh^2\left[\dfrac{\alpha_1\varepsilon_{2\|}\left(x-d_2\right)}{\varepsilon_1}\right]\right\}, & 0 < x < d_2, \end{cases} \tag{11}$$

we observe that it reaches its minima at the interfaces with vacuum $x = -d_1, x = d_2$ and it is exponentially peaked at the ENG-DPS interface $x = 0$, thereby yielding a *localized surface mode*.

The above features, markedly different from those exhibited by the SNG-DPS configurations considered in [7,8,18,20,21], closely resemble what observed in connection with matched ENG-MNG bi-layers studied in [3]. This suggests that the uniaxial DPS slab in our configuration may *effectively mimic* a MNG slab. Below, we explore to what extent this equivalence is fulfilled.

## C. Analogy between Uniaxial-DPS and MNG Media

For the assigned ENG slab parameters $\varepsilon_1$ and $d_1$, straightforward enforcement of the matching conditions in [3] yields the constitutive parameters

$$\begin{cases} \varepsilon_{2e} = \dfrac{d_1}{d_2}\left|\varepsilon_1\right|, \\ \mu_{2e} = \left(\dfrac{1}{\varepsilon_{2e}} - \dfrac{\varepsilon_{2e}}{\varepsilon_1^2}\right)\sin^2\theta_i - \dfrac{\varepsilon_{2e}}{\left|\varepsilon_1\right|}, \end{cases} \tag{12}$$

of an effective homogeneous, isotropic slab of thickness $d_2$ that would perfectly compensate the ENG slab for the incidence angle $\theta_i$. The medium described by the constitutive parameters in (12) is MNG for any incidence direction if $\varepsilon_{2e} \geq \left|\varepsilon_1\right|$ (i.e., $d_1 \geq d_2$). For $\varepsilon_{2e} < \left|\varepsilon_1\right|$ (i.e., $d_1 < d_2$), the MNG character is restricted to the incidence cone

$$\left|\sin\theta_i\right| \leq \varepsilon_{2e}\sqrt{\frac{\left|\varepsilon_1\right|}{\varepsilon_1^2 - \varepsilon_{2e}^2}}. \tag{13}$$

For such effective medium, let us consider the attenuation constant

$$\alpha_{2e} = k\sqrt{\sin^2\theta_i - \varepsilon_{2e}\mu_{2e}}, \tag{14}$$

and the transverse wave impedance

$$\eta_{2e} = \frac{i\eta\alpha_{2e}}{k\varepsilon_{2e}}, \tag{15}$$

with $\eta$ denoting the vacuum characteristic impedance, and compare them with the corresponding expressions for the uniaxial DPS medium, i.e., $\alpha_2$ in (5) and

$$\eta_2 = \frac{i\eta\alpha_2}{k\varepsilon_{2||}}. \tag{16}$$

It can readily be verified that the total-transmission conditions in (7) and (9) yield

$$\begin{cases} \varepsilon_{2||} = \varepsilon_{2e}, \\ \alpha_2 = \alpha_{2e}, \\ \eta_2 = \eta_{2e}. \end{cases} \tag{17}$$

In other words, the uniaxial DPS slab exhibits the *same* transverse-field distributions (and, hence, reflection and transmission responses) as the matched effective slab in (12), which may be MNG under appropriate conditions [see the discussion after (12)].

The above analogy closely resembles the one exploited in [24,25] in order to emulate a MNG medium via a metallic waveguide operating below cutoff under TM polarization. Both mechanisms rely on the *capacitive* character of transverse wave impedance of TM-polarized evanescent fields. While in [24,25] the underlying cutoff condition is generated by the metallic walls, in our configuration it is instead created by the *material* properties of a DPS uniaxial medium. This is in some ways consistent with the channeling properties of uniaxial metamaterials characterized by extreme material parameters [26]. Also in this case, strong spatial dispersion effects are hinted by the dependence of the effective permeability (12) on the incidence angle. Clearly, in view of this explicit dependence on the incidence angle, and the implicit dependence on frequency (given the inherently dispersive character of SNG media), we expect the above analogy to be practically realizable only under narrow-angle/frequency conditions (see below for numerical examples).

## 4. Representative Numerical Results

In order to elucidate the basic underlying phenomenology, we begin considering an ideal lossless configuration featuring a mildly opaque ENG slab with $\varepsilon_1 = -3$ and $d_1 = 0.1\lambda_0$, and an incidence direction $\theta_{i0} = 30°$; here and henceforth the subscript "$_0$" is used to identify the *fiducial* frequency/wavelength and incidence angle for which the tunneling conditions are *strictly* fulfilled. Among the infinite solutions of the total-transmission conditions in (7) and (9), we select the one with $d_2 = d_1$, which yields $\varepsilon_{2\perp} = 0.12$ and $\varepsilon_{2\|} = 3$. Figure 2 shows the transverse field (intensity and phase) distributions at resonance, from which the *complete tunneling* effect is clearly visible. As anticipated in Sec. 3.B, it can be observed that the effect is mediated by the excitation of a *localized surface mode* at the ENG-DPS interface (with evanescent field amplification in the ENG layer), and it does not imply any phase-delay accumulation. In other words, for the given polarization and incidence direction, the incident wavefront at the input interface $x = -d_1$ is perfectly reproduced at the output interface $x = d_2$, so that the bi-layer effectively behaves as an EM "nullity."

It is interesting to explore the sensitivity of the phenomenon with respect to the frequency, polarization, and incidence direction of the illumination, as well as to the unavoidable material dispersion and losses. To this aim, for the same configuration above, we consider a more realistic (dispersive, lossy) Drude-type model for the ENG medium,

$$\varepsilon_1(\omega)=1-\frac{\omega_{p1}^2}{\omega(\omega+i\gamma_1)}, \tag{18}$$

with the plasma angular frequency $\omega_{p1}$ and the damping coefficient $\gamma_1$ adjusted so that $\mathrm{Re}\left[\varepsilon_1(\omega_0)\right]\approx -3$ (with a loss-tangent $\sim 10^{-2}$ at resonance). For the uniaxial DPS slab, we assume a conventional mixing formula [27]

$$\begin{cases}\varepsilon_{2\perp}(\omega)=\left[\dfrac{\tau}{\varepsilon_a(\omega)}+\dfrac{1-\tau}{\varepsilon_b}\right]^{-1}, \\ \varepsilon_{2\|}(\omega)=\tau\varepsilon_a(\omega)+(1-\tau)\varepsilon_b,\end{cases} \tag{19}$$

typical of homogenized two-phase multilayer metamaterials, with the constituents modeled as

$$\varepsilon_a(\omega)=1-\frac{\omega_{pa}^2}{\omega(\omega+i\gamma_a)},\quad \varepsilon_b=4\left(1+10^{-3}i\right), \tag{20}$$

where the filling fraction $\tau$ and the other parameters are chosen so that $\mathrm{Re}\left[\varepsilon_{2\perp}(\omega_0)\right]\approx 0.12$ and $\mathrm{Re}\left[\varepsilon_{2\|}(\omega_0)\right]\approx 3$ (with a loss-tangent $\sim 10^{-3}$ at resonance). Figure 3 shows the corresponding transmittance response, for both TM and TE polarizations, as a function of frequency, from which a sharp, asymmetrical resonant line shape is observed at the nominal resonant frequency for the TM polarization, with a high-transmittance peak of ~93%. i.e., nearly a three-fold enhancement with respect to the typical transmittance level of the *standalone* ENG slab (also shown, as a reference, in the inset). The close-by zero-transmittance dip is attributable to a passage through zero of the (real part of) $\varepsilon_{2\perp}(\omega)$, for which the electric field at the slab entrance is required to be purely tangential, producing total reflection analogous to a perfect magnetic conductor for TM incidence [28]. Away from the resonance, as well as for the TE polarization (which does not exhibit any resonance), the transmittance levels are comparable with those typical of the standalone ENG slab. For the same configuration, Fig. 4 shows the angular response, from which a moderately broad transmittance peak centered at the fiducial incidence angle $\theta_i = \theta_{i0}$ is observed for the TM polarization. For increasing level of losses, a progressive decrease in the transmittance peak amplitude is observed (not shown here for brevity), qualitatively similar to what observed in other resonant tunneling phenomena [3].

In light of the analogy established in Sec. 3.C, it is insightful to compare the response of the uniaxial DPS slab above with that of an effective "matched" (according to [3]) homogeneous, isotropic MNG slab of same thickness featuring

$$\varepsilon_{2e} = 3, \quad \mu_{2e}(\omega_0) = -1. \tag{21}$$

where we assume a Drude-type frequency dispersion for the permeability, analogous to (20). Substituting, in the bi-layer, the uniaxial DPS slab with such MNG slab, would yield, at the fiducial frequency and incidence angle, the same tunneling condition, with field distributions identical to those in Fig. 3. Figures 5 and 6 compare the reflection coefficient (magnitude and phase) responses of the *standalone* (uniaxial DPS and MNG) slabs, as a function of frequency and angle, respectively. Besides the expected *perfect match* of the responses at the fiducial frequency and incidence angle, it can be observed that the agreement rapidly deteriorates within a relatively small neighborhood. Qualitatively similar results (not shown here for brevity) are observed for the transmission coefficient. Therefore, we may conclude that the uniaxial DPS slab may effectively mimic a MNG-type response only within narrow frequency/angular ranges.

As a further example, we consider a more critical configuration, featuring an ENG slab with increased opacity ($\mathrm{Re}\left[\varepsilon_1\left(\omega_0\right)\right]=-100$ and $d_1=\lambda_0/100$), and a closer-to-normal nominal incidence direction $\theta_{i0}=15°$. From the total-transmission conditions in (7) and (9), selecting $d_2=100d_1/3=\lambda_0/3$, we obtain $\mathrm{Re}\left[\varepsilon_{2\perp}\left(\omega_0\right)\right]=0.065$ and $\mathrm{Re}\left[\varepsilon_{2\|}\left(\omega_0\right)\right]=3$, i.e., a rather *extreme* anisotropy. Figures 7 and 8 show the corresponding frequency and angular responses, respectively. By comparison with the previous example (Figs. 3 and 4), qualitatively similar considerations hold, with an increased frequency and angular selectivity [29]. The high-transmittance peak (nearly 70%) turns out to be moderately lower in absolute terms, but considerably higher in terms of enhancement (nearly a factor eight) with respect to the standalone ENG slab.

Finally, in order to include retardation effects, we consider a configuration featuring *electrically thicker* slabs, namely, $\mathrm{Re}\left[\varepsilon_1\left(\omega_0\right)\right]=-1.5$, $\mathrm{Re}\left[\varepsilon_{2\perp}\left(\omega_0\right)\right]=0.115$, $\mathrm{Re}\left[\varepsilon_{2\|}\left(\omega_0\right)\right]=1.5$, and $d_1=d_2=\lambda_0/2$, for $\theta_{i0}=30°$ and a reduced level of losses. From the corresponding frequency and angular responses, shown in Figs. 9 and 10, respectively, we observe that the tunneling effect is still possible, although the resonant peaks are much narrower than the previous examples, and the peak transmittance is much more sensitive to the level of losses. This is expected, due to the larger quality-factor of the resonant tunneling in this example, and is consistent with similar sensitivity observed in thicker ENG-MNG matched bi-layers [3].

## 5. Conclusions

In this paper, we have studied an interesting EM tunneling effect that can take place in bi-layers featuring a homogeneous, isotropic ENG slab paired with a homogenous, anisotropic (uniaxial) DPS slab, under appropriate TM-polarized oblique-incidence. After a rigorous analytical derivation of the total-transmission conditions (for the ideal lossless case), we have elucidated the underlying physical mechanisms, and emphasized the strong analogies with the ENG-MNG matched pairs in [3]. Moreover, we have carried out a parametric study aimed at exploring the sensitivity of the phenomenon with respect to frequency, polarization and incidence direction of the illumination, also taking into account material dispersion and losses.

The possibility of obtaining, for only one polarization, high-transmittance peaks with strong frequency/angular sensitivity (see, e.g., Figs. 7-10) may find application to polarizing frequency/spatial filters or beam splitters. Our results also indicate the possibility of emulating, within narrow frequency/angular ranges, the response of an MNG slab via an arguably simpler to realize uniaxial DPS slab.

In connection with possible extensions/generalization, we note that the results pertaining to a configuration featuring an MNG slab paired with a uniaxial *magnetic* DPS slab, under TE-polarized illumination, follow straightforwardly from duality considerations. Finally, it is worth emphasizing that, paralleling the approaches in [30,31], our results may be generalized to tunnel barriers of different nature (e.g., quantum-mechanical).

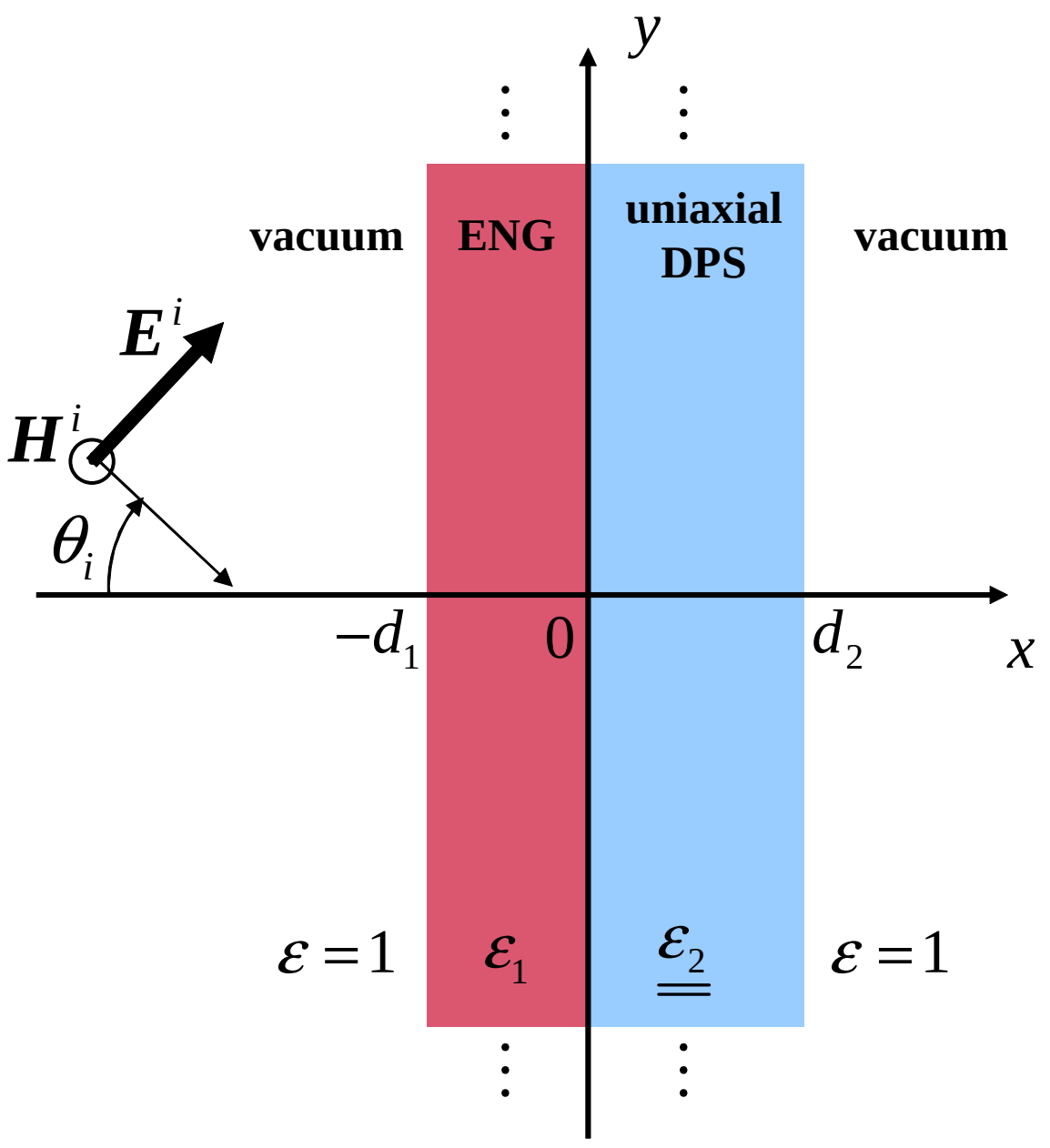

Fig. 1. (Color online) Problem schematic in the associated Cartesian reference system: We consider a homogeneous, isotropic slab of ENG material of thickness $d_1$ and relative permittivity $\varepsilon_1$ (with $\mathrm{Re}(\varepsilon_1) < 0$) paired with a homogeneous, anisotropic (uniaxial) DPS slab of thickness $d_2$ and relative permittivity tensor $\underline{\underline{\varepsilon}}_2$ given in (1). The bi-layer is immersed in vacuum, and is illuminated by an obliquely-incident, TM-polarized plane wave.

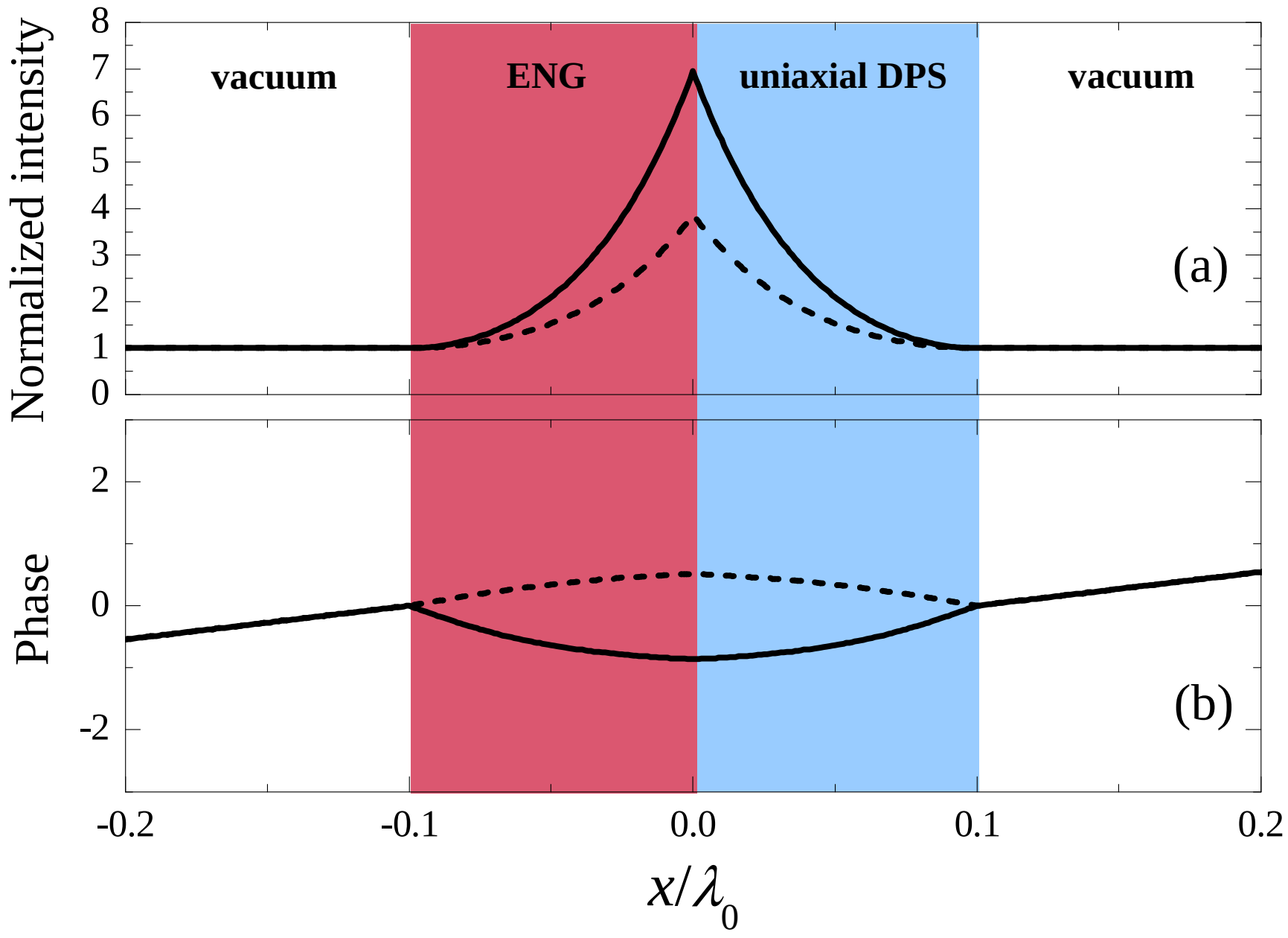

Fig. 2. (Color online) Intensity (a) and phase (b) distributions of transverse magnetic (solid) and electric (dashed) fields (normalized by the incident values) at resonance, for an ideal, lossless bi-layer as in Fig. 1, with $\varepsilon_1 = -3$, $d_1 = d_2 = 0.1\lambda_0$, $\varepsilon_{2\perp} = 0.12$, and $\varepsilon_{2\parallel} = 3$, under TM-polarized, obliquely-incident illumination with $\theta_i = \theta_{i0} = 30°$.

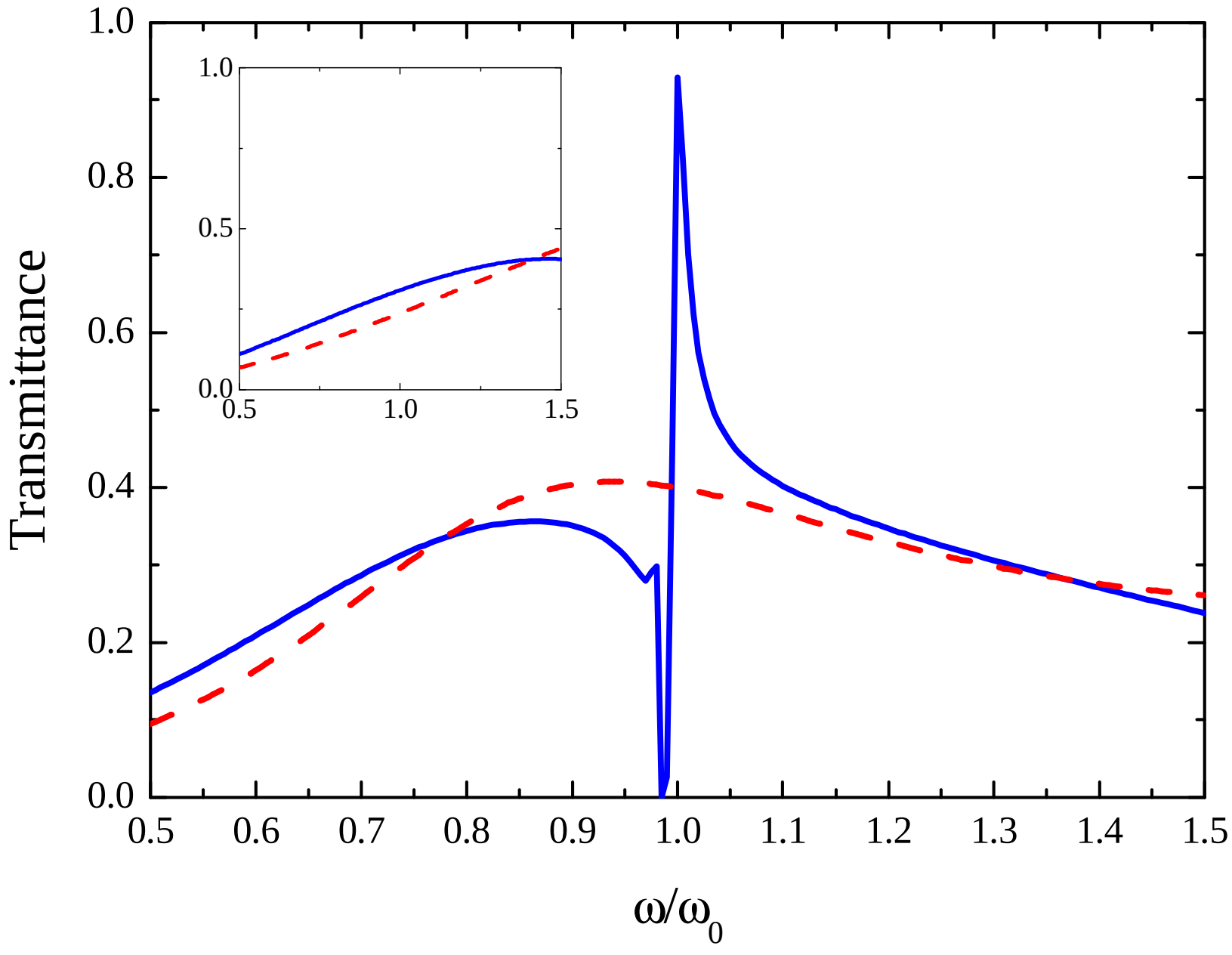

Fig. 3. (Color online) Transmittance (for $\theta_i = \theta_{i0} = 30°$) frequency response, for TM (blue-solid) and TE (red-dashed), pertaining to the parameter configuration as in Fig. 2, but considering for the ENG medium the Drude-type model in (18) with $\omega_{p1} = 2\omega_0$, $\gamma_1 = 3.75 \cdot 10^{-3} \omega_{p1}$ (so that $\mathrm{Re}\left[\varepsilon_1(\omega_0)\right] \approx -3$), and for the uniaxial DPS medium the mixing rules in (19) and (20), with $\tau = 0.252$, $\omega_{pa} = 0.984\omega_0$, $\gamma_a = 3.24 \cdot 10^{-4} \omega_{pa}$ (so that $\mathrm{Re}\left[\varepsilon_{2\perp}(\omega_0)\right] \approx 0.12$ and $\mathrm{Re}\left[\varepsilon_{2\|}(\omega_0)\right] \approx 3$). Also shown as a reference (in the inset) is the response of the *standalone* ENG slab.

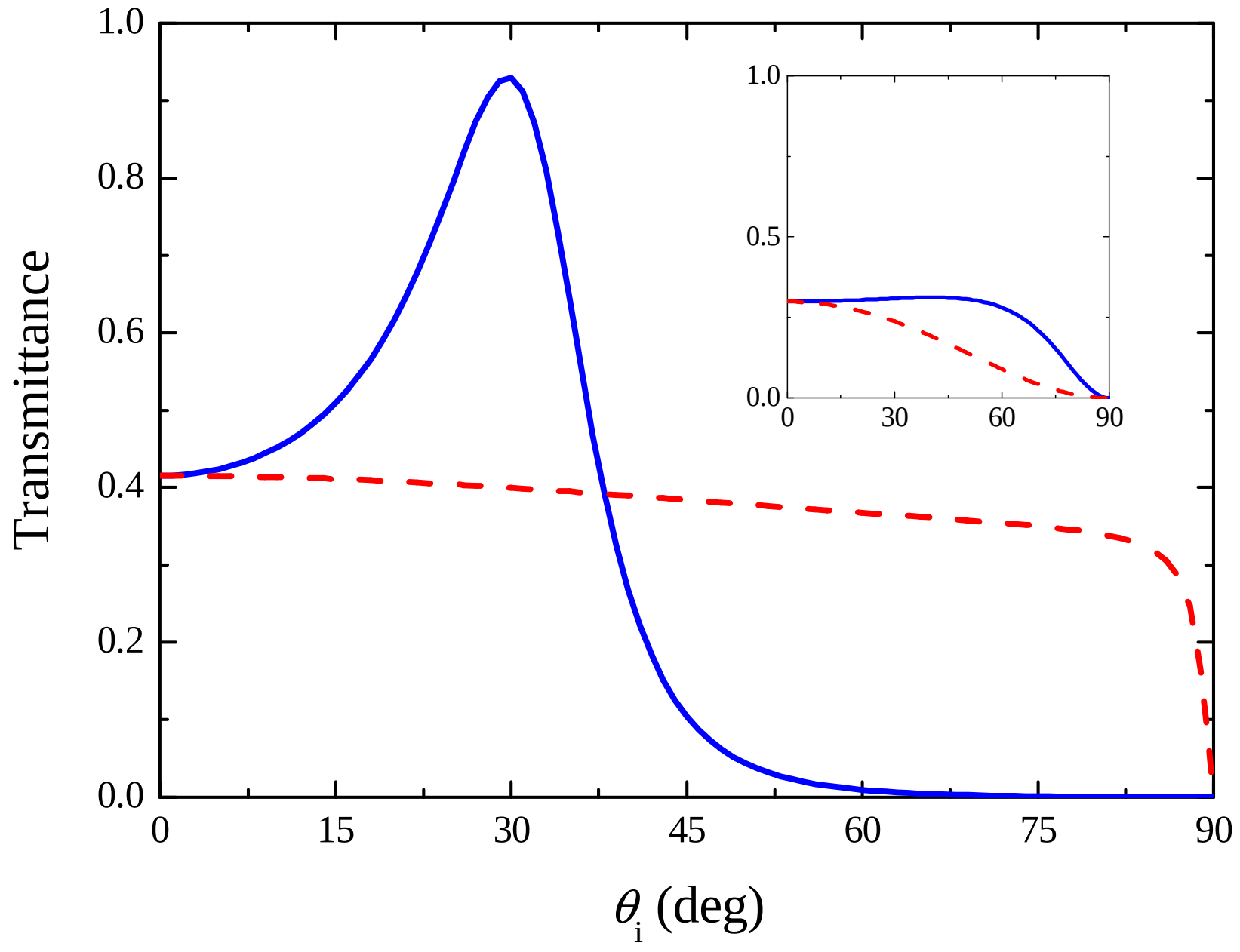

Fig. 4. (Color online) As in Fig. 3, but angular response at resonance.

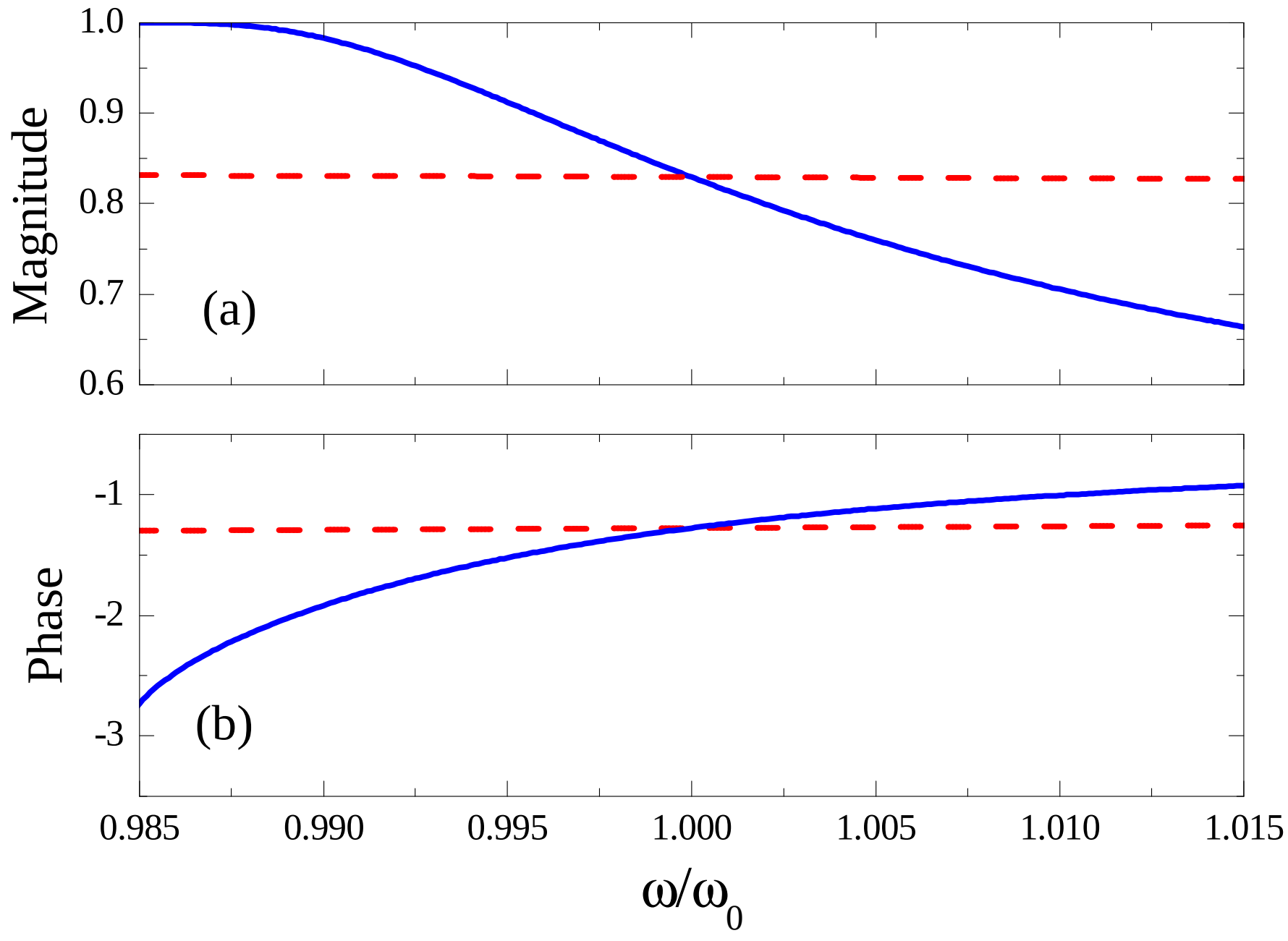

Fig. 5. (Color online) Magnitude (a) and phase (b) of the reflection-coefficient frequency response (for $\theta_i = \theta_{i0} = 30°$) pertaining to the *standalone* uniaxial DPS slab (blue-solid) with parameters as in Figs. 2 and 3 (for TM polarization, and assuming zero losses), compared with that of an effective (homogeneous, isotropic) matched (cf. (21)) MNG slab (red-dashed), with the relative permeability described by a lossless Drude-type model $\mu_{2e}(\omega) = 1 - 2\omega_0^2/\omega^2$ .

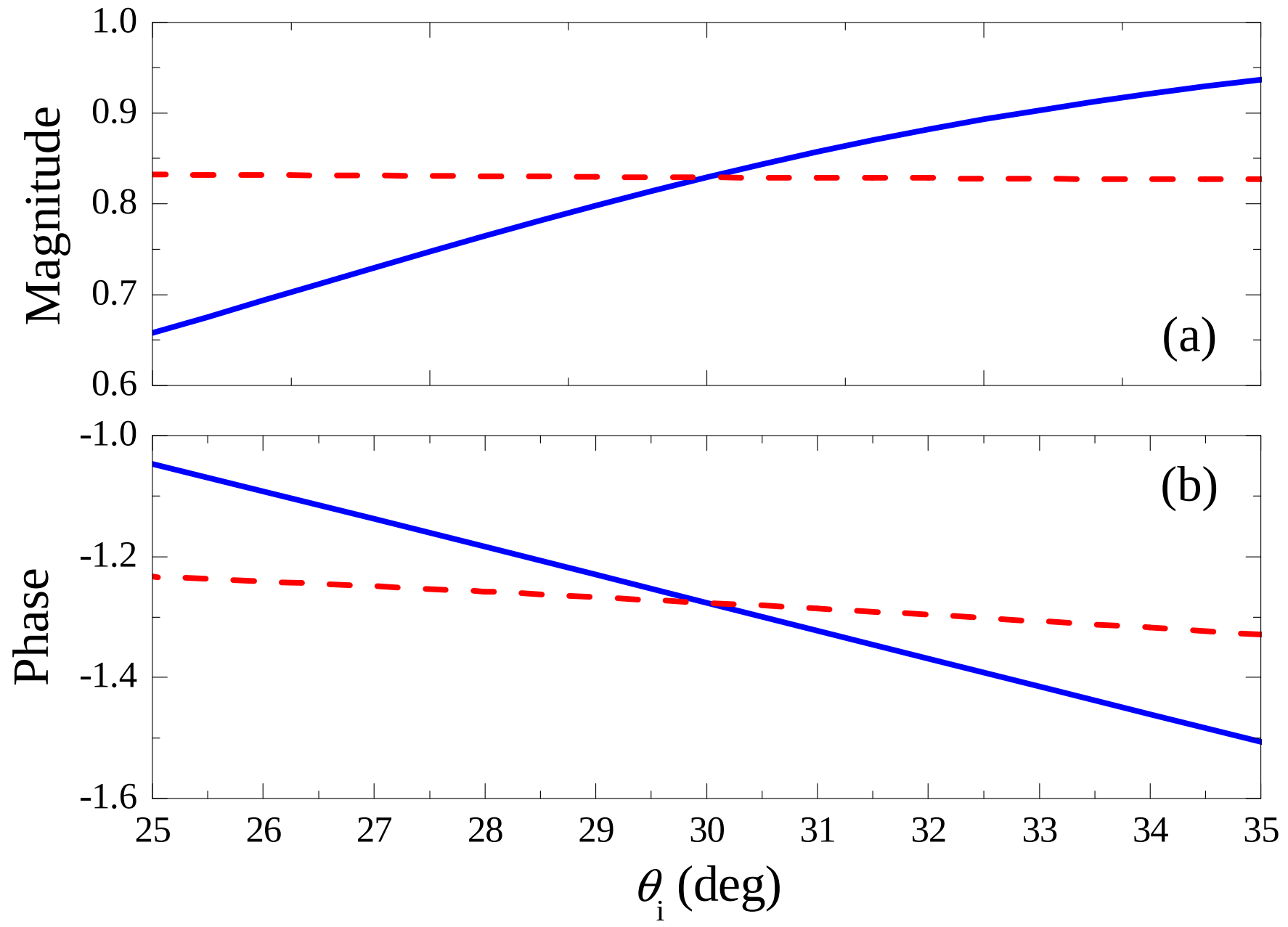

Fig. 6. (Color online) As in Fig. 5, but angular response at resonance.

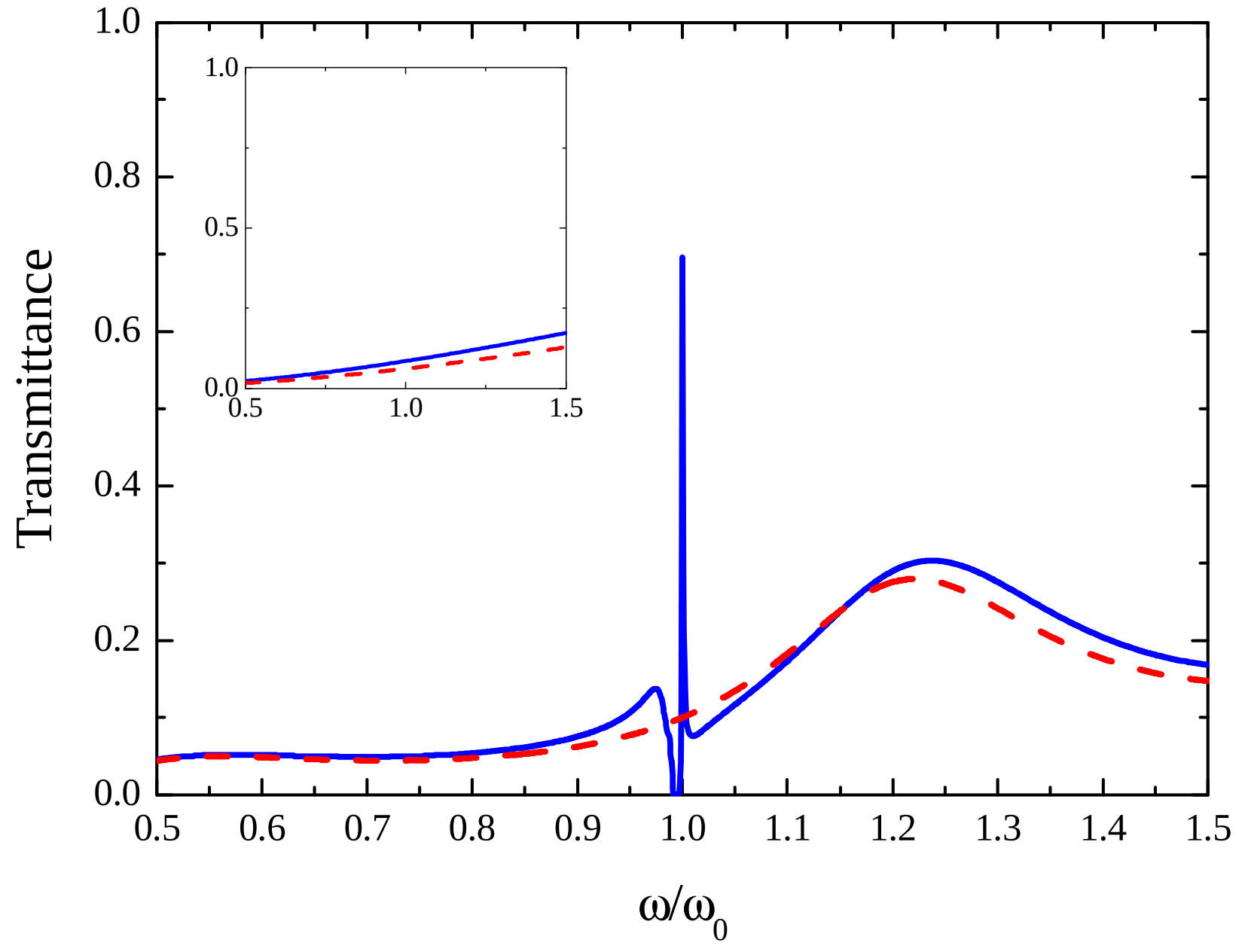

Fig. 7. (Color online) As in Fig. 3, but for $\theta_i = \theta_{i0} = 15°$, and an ENG slab (cf. (18)) with $\omega_{p1} = 10.05\omega_0$, $\gamma_1 = 9.85 \cdot 10^{-4}\omega_{p1}$ (i.e., $\mathrm{Re}\left[\varepsilon_1(\omega_0)\right] \approx -100$), and $d_1 = \lambda_0/100$, and a matched uniaxial DPS slab (cf. (19) and (20)) with $\tau = 0.251$, $\omega_{pa} = 0.992\omega_0$, $\gamma_a = 1.69 \cdot 10^{-3}\omega_{pa}$ (i.e., $\mathrm{Re}\left[\varepsilon_{2\perp}(\omega_0)\right] \approx 0.065$ and $\mathrm{Re}\left[\varepsilon_{2\|}(\omega_0)\right] \approx 3$), and $d_2 = \lambda_0/3$.

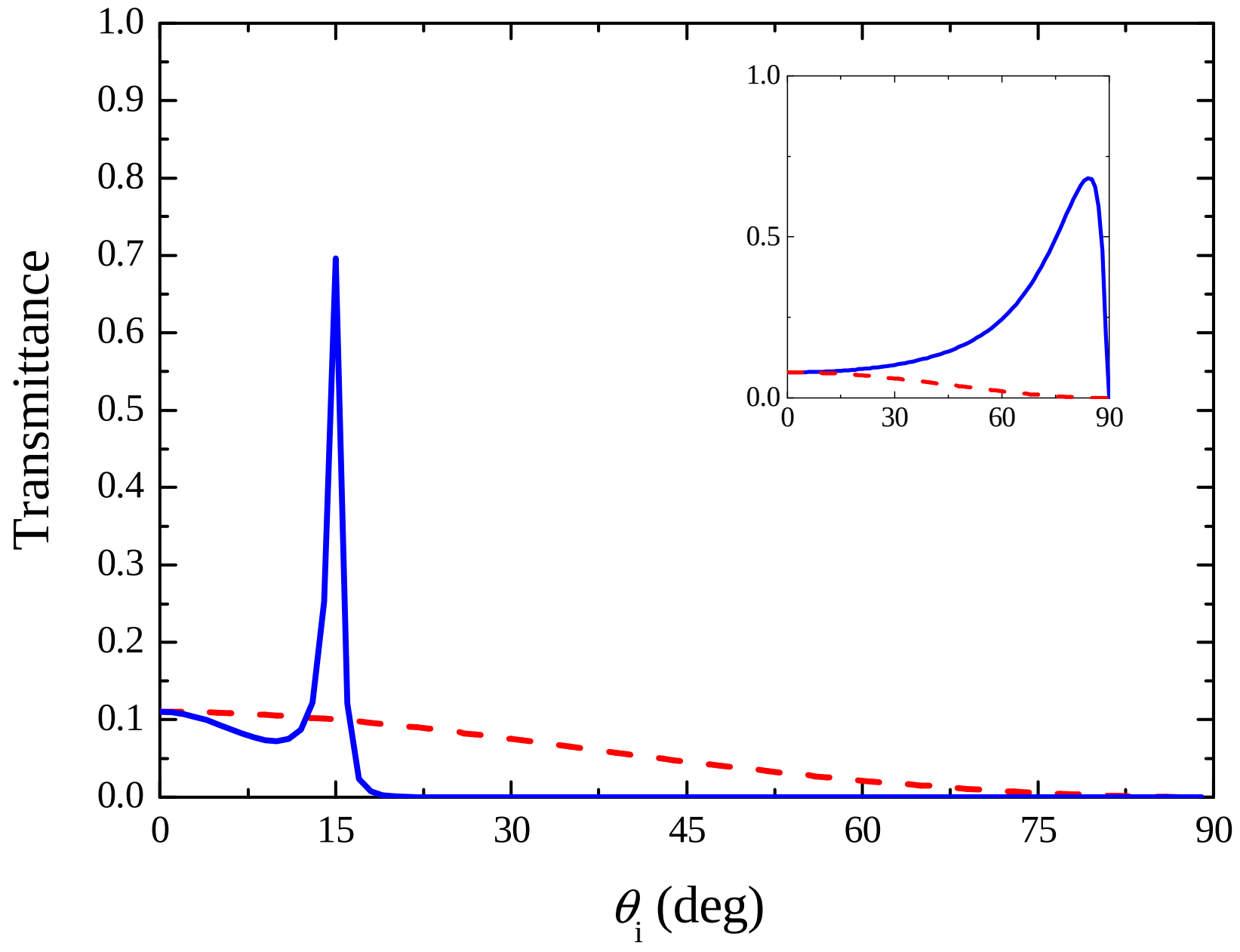

Fig. 8. (Color online) As in Fig. 7, but angular response at resonance.

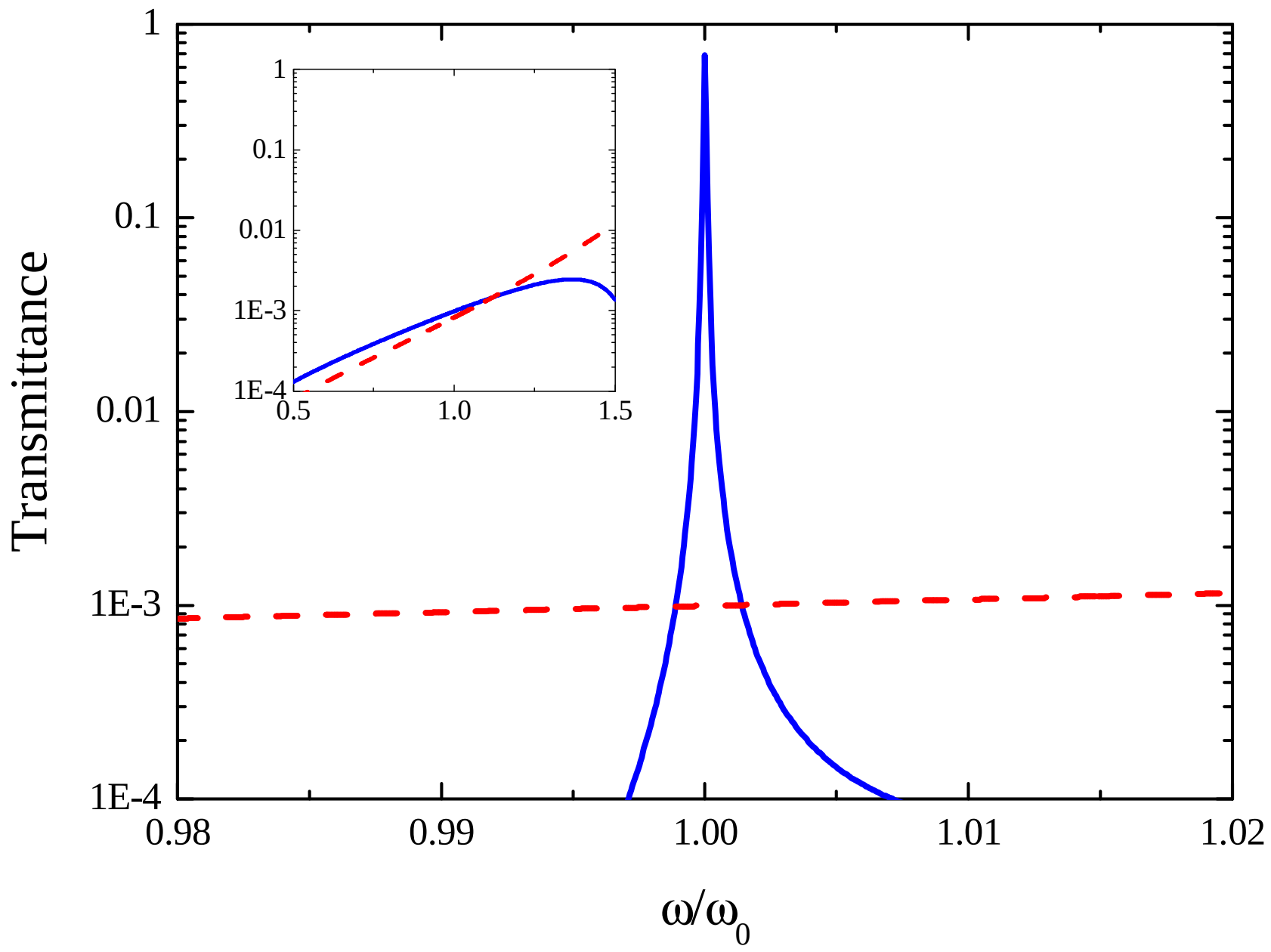

Fig. 9. (Color online) As in Fig. 3, but for an ENG slab (cf. (18)) with $\omega_{p1} = 1.58\omega_0$, $\gamma_1 = 3.8\cdot 10^{-5}\omega_{p1}$ (i.e., $\mathrm{Re}\left[\varepsilon_1(\omega_0)\right] \approx -1.5$), and $d_1 = \lambda_0/2$, and a matched uniaxial DPS slab (cf. (19) and (20)) with $\tau = 0.637$, $\omega_{pa} = 0.962\omega_0$, $\gamma_a = 8.34\cdot 10^{-6}\omega_{pa}$ (i.e., $\mathrm{Re}\left[\varepsilon_{2\perp}(\omega_0)\right] \approx 0.115$ and $\mathrm{Re}\left[\varepsilon_{2\parallel}(\omega_0)\right] \approx 1.5$), $\varepsilon_b = 4\left(1+10^{-4}i\right)$, and $d_2 = \lambda_0/2$. Note the semi-log scale and the narrower frequency range considered.

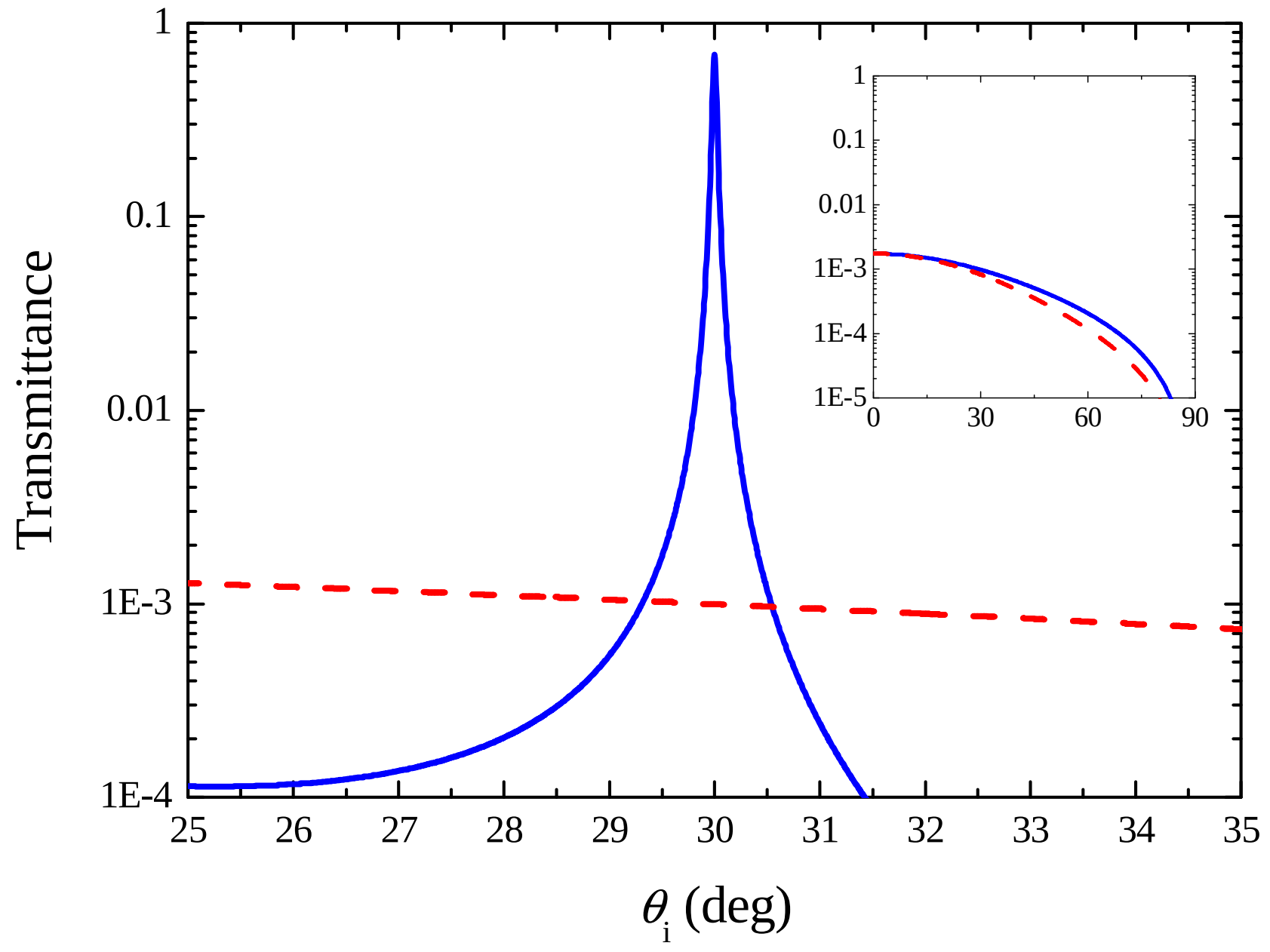

Fig. 10. (Color online) As in Fig. 9, but angular response at resonance. Note the semi-log scale and the narrower angular range considered.